\documentclass[prb,twocolumn,aps,amsmath,amssymb]{revtex4}
\usepackage[T1]{fontenc}
\usepackage[utf8]{inputenc}
\usepackage{graphicx}
\usepackage{amsmath}
\usepackage{amssymb}
\usepackage{dcolumn}
\usepackage{float}
\usepackage{bm}
\usepackage{amsfonts,times}
\usepackage{xcolor}
\usepackage[toc,page]{appendix}
\usepackage[breaklinks=true,colorlinks,citecolor=blue,linkcolor=blue,urlcolor=blue]{hyperref}

\DeclareMathAlphabet{\bi}{OML}{cmm}{b}{it}

\def\be{\begin{equation}}
\def\ee{\end{equation}}
\def\bearr{\begin{eqnarray}}
\def\eearr{\end{eqnarray}}

\begin{document}
\title{Effect of magnetic field on the electronic properties of an $\alpha$-$T_3$ ring}
\author{Mijanur Islam$^1$}
\author{Tutul Biswas$^{2\footnote{Corresponding author}}$}
\email{tbiswas@nbu.ac.in}
\author{Saurabh Basu$^1$}

\affiliation{$^1$ Department of Physics, Indian Institute of Technology-Guwahati, Guwahati-781039, India.}
\affiliation{$^2$ Department of Physics, University of North Bengal, Raja Rammohanpur-734013, India.}
\normalsize

\begin{abstract}
We consider a quantum ring of a certain radius $R$ built from a sheet of the $\alpha$-$T_3$ lattice and solve for its spectral properties in the presence of an external magnetic field. The energy spectrum consists of a conduction band, a valence band, and a zero-energy flat band, all having a number of discrete levels which can be characterized by the angular momentum quantum number $m$. The energy levels in the flat band are infinitely degenerate irrespective of the value of $\alpha$. We reveal a twofold degeneracy of the levels in the conduction band as well as in the valence band for 
$\alpha=0$ and $\alpha=1$. However, the $m=0$ level for $\alpha=1$ is an exception.  Corresponding to an intermediate value of $\alpha$, namely, $0<\alpha<1$, the energy levels become nondegenerate. The scenario for the degeneracy of the energy levels remains unaltered when the ring is threaded by a magnetic flux which is an integer multiple of the flux quantum. We comment on the energy levels which are relevant for low-energy physics by studying their radius dependence in the presence of a magnetic field.
We also calculate the persistent current, which exhibits quantum oscillations as a function of the magnetic field with a period of one flux quantum at a particular Dirac point, which is often referred to as a valley. The total persistent current comprising the contributions from both the valleys is zero in the cases corresponding to $\alpha=0$ and
$\alpha=1$. However, the total current oscillates with a periodicity of one flux quantum for any intermediate value of $\alpha$. 
We also explore the effect of a mass term (that breaks the sublattice symmetry) in the Hamiltonian. In the absence of a magnetic field, the energy levels in the flat band become dispersive, except for the $m=0$ level in the case of $\alpha=1$. In the presence of the field, each of the flat band levels becomes dispersive for any $\alpha \neq0$.  Finally, we also see the effect of the mass term on the behavior of the persistent current, which shows a periodicity of one flux quantum, but the total current remains finite for all values of $\alpha$.
 
\end{abstract}

\maketitle
\section{Introduction}
Electronic properties at low dimensions with various geometries have continued to fascinate the scientific community over the years. Among such structures quantum rings (QRs) are widely celebrated due to their peculiar 
electronic properties. The fabrication of nanoscale quantum rings[\onlinecite{Fabri_Ring1,Fabri_Ring2}] in semiconductor heterostructures has aided in the understanding of the theoretical results on the subject[\onlinecite{Th_Ring1,Th_Ring2}]. A QR can host persistent current[\onlinecite{Buttiker}] when it is threaded by a magnetic flux. This persistent current is closely related to the Aharonov-Bohm (AB) effect[\onlinecite{AB_Eff}]. A reasonable number of studies[\onlinecite{Pers1, Pers2, Pers3, Pers4, Pers5, Pers6, Pers7, Pers8, Pers9}] has been devoted to confirming the existence of the persistent current in ringlike quantum structures. The consideration of spin-orbit interaction of mainly the Rashba type[\onlinecite{Sem_Ring1}] has given rise to various spectacular spin-related phenomena[\onlinecite{Spin_dep1, Spin_dep2, Spin_dep3, Spin_dep4, Spin_dep5, Spin_dep6}] in semiconductor QRs.  

With the advent of graphene[\onlinecite{Graph_exp1, Graph_exp2, Graph_exp3, Graph_exp4}], there has been immense interest in the different nanostructures, including QRs based on it. QRs fabricated by lithographic techniques[\onlinecite{Grap_Lith1,  Grap_Lith2,Grap_Lith3}] provide suitable platforms to study the AB effect experimentally. There are a number of numerical[\onlinecite{Graph_Numr1, Graph_Numr2, Graph_Numr3, Graph_Numr4, Graph_Numr5, Graph_Numr6, Graph_Numr7, Graph_Numr8, Graph_Numr9, Graph_Numr10, Graph_Numr11}] and analytical[\onlinecite{Graph_Model1, Graph_Model2, Graph_Bilayer, Graph_Spin, hybrid_grapR}] studies on graphene QRs dealing with both charge and spin persistent currents, valley splitting, etc. It has been demonstrated that graphene QRs have potential applications in future optoelectronic[\onlinecite{Graph_Opto}] and interferometric[\onlinecite{Graph_Interf}] devices.
 
On the other hand, an interesting variant of the honeycomb structure of graphene with $T_3$ symmetry, usually known as the dice lattice, exists[\onlinecite{Dice_1, Dice_2}]. Here, the honeycomb lattice is augmented by an extra site located at the center of each hexagon. Three inequivalent sites in a unit cell effectively introduce an enlarged pseudospin $S=1$. 
It has been argued that a particular arrangement of three counterpropagating pairs of laser beams  can produce an optical dice lattice[\onlinecite{Dice_Opt}] in the cold atomic environment.
It has further been proposed that a dice lattice can be fabricated in a heterostructure of cubic lattices, namely, SrTiO$_3$/SrIrO$_3$/SrTiO$_3$[\onlinecite{Dice_Real}]. A more generalized lattice, called the $\alpha$-$T_3$ lattice[\onlinecite{Illes_Thesis}], demonstrates a smooth changeover with the variation of the parameter $\alpha$ from graphene $(\alpha=0)$ to the dice lattice $(\alpha=1)$. The electronic dispersion of the $\alpha$-$T_3$ lattice with $\alpha=1/\sqrt{3}$ can be realized in a Hg$_{1-x}$Cd$_x$Te quantum well corresponding to a certain critical doping
[\onlinecite{alp_T3_real}]. Within the nearest-neighbor tight-binding 
framework, the low-energy spectrum of the $\alpha$-$T_3$ lattice near a particular valley, governed by the Dirac-Weyl Hamiltonian with an enlarged pseudospin ($S>1/2$), consists of three bands, with two dispersive bands which are linear in momentum and a zero-energy flat band. 
It is well known that the Berry phase in the $\alpha$-$T_3$ lattice is a function of 
$\alpha$. This variable Berry phase further causes the magnetization to exhibit a smooth crossover from a diamagnetic ($\alpha=0$) to a paramagnetic ($\alpha=1$) behavior across the critical value of $\alpha$, namely,
$\alpha_c=0.495$[\onlinecite{aT3_Para_Dia}]. A plethora of studies were performed in recent years to probe various equilibrium[\onlinecite{T3_Hall1, T3_Hall2, Klein1, Klein2, Weiss, ZB, Plasmon1, Plasmon2, Plasmon3, Plasmon4, Mag_Opt1, Mag_Opt2, Mag_Opt3, Mag_Opt4, RKKY1, RKKY2, Min_Con, Ghosh_Topo, spin_hall, aT3_Mijanur}] and nonequilibrium[\onlinecite{Bashab1, Bashab2, Iurov_Floq, Mojarro, TB_Floq, Trans_aT3, TB_Trans}] properties of the $\alpha$-$T_3$ lattice. 
 
 \begin{figure}[h!]
\centering
\includegraphics[width=8.5cm, height=7.0cm]{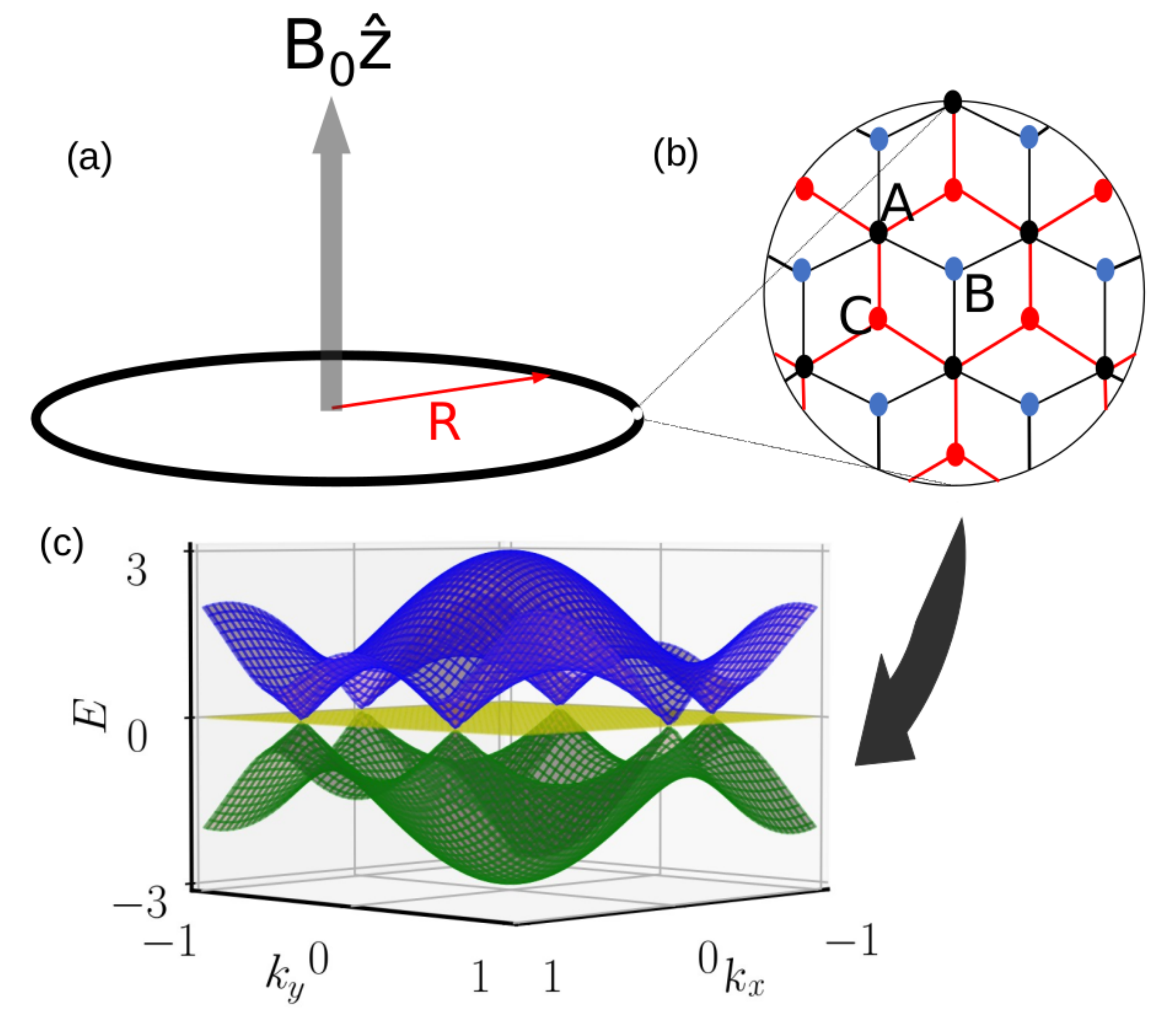}
\caption{(a) A schematic diagram of the $\alpha$-$T_3$ ring of radius $R$ subjected to a transverse magnetic field $\bm B=B_0\hat{z}$. (b) The structure of the $\alpha$-$T_3$ lattice is depicted in the zoomed portion. $A$, $B$, and $C$ lattice sites are shown by black, blue, and red dots, respectively. (c) The zero-field bulk band structure consists of dispersive conduction and valence bands and a nondispersive zero-energy band. The conduction band touches the valence band at the Dirac points, known as valleys, in the first Brillouin zone. Around those points, the spectrum becomes linearly dispersive. Here, $E$, $k_x$, and $k_y$ are in arbitrary units.}
\label{fig:ring_geo}
\end{figure}

To the best of our knowledge, no attention has been paid to a QR made in the 
$\alpha$-$T_3$ lattice, which we call an $\alpha$-$T_3$ ring. Therefore, it would be interesting to study the electronic properties of the $\alpha$-$T_3$ ring. Particularly, we intend to see how the energy spectrum evolves as we migrate from graphene ($\alpha=0$) to the dice lattice ($\alpha=1$). The inclusion of an external magnetic field would facilitate studies on the evolution of the spectral properties, the persistent current, and the interplay of the AB effect therein. With such a motivation, we consider an $\alpha$-$T_3$ ring in the presence of a magnetic field and study various properties as a function of $\alpha$, which is a parameter in this work. Finally, we also study the effect of a mass term (which is different for different sublattices) on the spectral properties and the corresponding persistent current.

The rest of this paper is organized in the following way. In Sec. \ref{Mod} we discuss various properties of the $\alpha$-$T_3$ ring, including the energy spectrum and the persistent current. In Sec. \ref{Sec_Mass}, we discuss the effect of a mass term on the spectrum as well as on the persistent current of the ring. We summarize our results in Sec. \ref{Sum}.

\section{The $\alpha$-$T_3$ ring}
\label{Mod}
We consider a ring of radius $R$, in the $x$-$y$ plane, made from the 
$\alpha$-$T_3$ lattice, as depicted in Fig.\,\ref{fig:ring_geo}(a). The geometric structure of the 
$\alpha$-$T_3$ lattice is shown in Fig. \ref{fig:ring_geo}(b). A unit cell contains three inequivalent lattice sites, namely, $A$, $B$, and $C$. Sites $A$ and $B$ form the honeycomb structure (graphene) with the nearest-neighbor hopping parameter $t$. The center site $C$ is connected to only three surrounding $A$ sites with hopping $\alpha t$, where $0<\alpha\leq1$.  

Additionally, the ring is also subjected to a perpendicular magnetic field $\bm{B} = B_0 \hat{z}$, where $B_0$ is a constant. Before discussing the details of the $\alpha$-$T_3$ ring, we briefly talk about the bulk band structure of the $\alpha$-$T_3$ lattice.
 
In the vicinity of a particular valley ($K$ or $K^\prime$) characterized by the index $\zeta=\pm1$, the low-energy excitations in the $\alpha$-$T_3$ lattice can be described by the following Dirac-Weyl Hamiltonian:
\begin{eqnarray}\label{HamT3}
H^\zeta=v_F(\zeta\pi_xS_x+\pi_yS_y),
\end{eqnarray}
where $v_F$ is the Fermi velocity and $\pi_x$ and $\pi_y$ are the components of the canonical momentum operator defined via ${\bm \pi}={\bm p}+e{\bm A}$, where $\bm p$ denotes the in-plane mechanical momentum operator and $\bm A$ is the vector potential. Here, the $x$ and $y$ components of the pseudospin operator $\bm S$ associated with the $\alpha$-$T_3$ lattice are given by
\begin{equation*}
S_x = \begin{pmatrix}
0 & \cos\phi & 0\\
\cos\phi & 0 & \sin\phi\\
0 & \sin\phi & 0
\end{pmatrix}
\end{equation*}
and
\begin{equation*}
S_y =\begin{pmatrix}
0 & -i\cos\phi & 0\\
i\cos\phi & 0 & -i\sin\phi\\
0 & i\sin\phi & 0
\end{pmatrix},
\end{equation*}
respectively, with $\tan\phi=\alpha$. The $z$ component of ${\bm S}$ can be directly obtained from the commutation relation $[S_x,S_y]=iS_z$. 
The low-energy zero-field bulk band structure of the $\alpha$-$T_3$ lattice consists of two linearly dispersive bands: $E_k^\pm=\pm\hbar v_Fk$ and a 
zero-energy flat band: $E=0$, as shown in Fig. \ref{fig:ring_geo}(c).

In general, the vector potential $\bm A$ corresponding to a uniform magnetic field $\bm B$ is given by ${\bm A}=\frac{1}{2}({\bm B} \times {\bm r})$, where $\bm r$ is the position vector. Since $\bm B=B_0\hat{z}$, we choose $\bm A$ in the symmetric gauge to be 
$\bm A=B_0(-y\hat{x}+x\hat{y})/2$. For our problem it the more convenient to write $\bm A$ in the polar coordinates $(r, \theta)$ as 
$\bm A=\frac{1}{2}rB_0\hat{\theta}$.

Therefore, the Hamiltonian in Eq. (\ref{HamT3}) can be expressed in polar coordinates as
\begin{widetext}
\begin{eqnarray}
\label{Pol_Ham}
H^\zeta=\hbar v_F \begin{pmatrix}
0 & \cos\phi\,e^{-i\zeta\theta}\big(-i\zeta\frac{\partial}{\partial r}-\frac{1}{r}\frac{\partial}{\partial \theta}-\frac{ieB_0r}{2\hbar}\big) & 0\\
\cos\phi\,e^{i\zeta\theta}\big(-i\zeta\frac{\partial}{\partial r}+\frac{1}{r}\frac{\partial}{\partial \theta}+\frac{ieB_0r}{2\hbar}\big) & 0 & \sin\phi\,e^{-i\zeta\theta}\big(-i\zeta\frac{\partial}{\partial r}-\frac{1}{r}\frac{\partial}{\partial \theta}-\frac{ieB_0r}{2\hbar}\big)\\
0 & \sin\phi\,e^{i\zeta\theta}\big(-i\zeta\frac{\partial}{\partial r}+\frac{1}{r}\frac{\partial}{\partial \theta}+\frac{ieB_0r}{2\hbar}\big) & 0
\end{pmatrix}.
\end{eqnarray}
\end{widetext} 
The  eigenstates of $H^\zeta$ can be obtained as
\begin{eqnarray}\label{EigT3}
\psi^{m\zeta}(r,\theta)=\begin{pmatrix}
\chi_1(r)e^{i(m-\zeta)\theta}\\
\chi_2(r)e^{im\theta}\\
\chi_3(r)e^{i(m+\zeta)\theta}
\end{pmatrix},
\end{eqnarray}
where the integer $m$ labels the orbital angular momentum quantum number and
$\chi_i$ denotes the amplitudes corresponding to the three sublattices. Note that the Hamiltonian $H^\zeta$ commutes with the $z$ component of the total angular momentum operator defined as $J_z=L_z+S_z$, where $L_z=-i\hbar \frac{\partial}{\partial \theta}$ is the orbital angular momentum operator and $S_z$ is the pseudospin operator. 
Therefore, Eq. (\ref{EigT3}) is also the eigenstate of $J_z$. 

Now, we consider a strictly one-dimensional(1D) ring of radius $R$ such that the radial part is frozen in the eigensolution[\onlinecite{Sem_Ring1, Spin_dep2, Spin_dep5,Graph_Numr8, Graph_Model1, Graph_Model2}]. For the sake of the Hermiticity of the Hamiltonian in ring geometry one should make the replacements 
$r\rightarrow R$ and $\frac{\partial}{\partial r}\rightarrow-\frac{1}{2R}$. These replacements are also obvious because the radial momentum vanishes in the strict 1D limit (see the Appendix A for details). Therefore, the Hamiltonian corresponding to an ideal $\alpha$-$T_3$ ring is given by
\begin{widetext}
\begin{equation}
H^\zeta_{\rm ring}=\frac{\hbar v_F}{R}\begin{pmatrix}
0 & -i(m+\beta-\frac{\zeta}{2})\cos\phi & 0\\
i(m+\beta-\frac{\zeta}{2})\cos\phi & 0 & -i(m+\beta+\frac{\zeta}{2})\sin\phi\\
0 & i(m+\beta+\frac{\zeta}{2})\sin\phi & 0
\end{pmatrix},
\end{equation}
\end{widetext}
with $\beta=\Phi/\Phi_0$, where $\Phi=\pi R^2 B_0$ is a magnetic flux through the ring and 
$\Phi_0$ is the usual flux quantum.
We obtain the energy spectrum as
\begin{eqnarray}
\label{Energ_Eig}
E_{\rm FL}^{m\zeta}=0,~~~~~~ E_\pm^{m\zeta} =\pm \frac{\hbar v_F}{R}
\Delta_m^\zeta (\alpha,\beta).
\end{eqnarray}
Here, $\Delta_m^\zeta (\alpha,\beta)$ is defined as
\begin{eqnarray}\label{Delm}
\Delta_m^\zeta (\alpha,\beta)=\sqrt{\big(m+\beta)^2+\frac{1}{4}-\zeta\big(m+\beta\big)\frac{1-\alpha^2}{1+\alpha^2}}.
\end{eqnarray}
The energy spectrum in Eq. (\ref{Energ_Eig}) for the $\alpha$-$T_3$ ring consists of a zero-energy flat band $E_{\rm FL}$ alongside a number of discrete levels in the conduction band ($E_{+}$) and the valence band ($E_{-}$).

The normalized wave functions corresponding to $E_\pm^{m\zeta}$ in a particular valley are
obtained as
\begin{eqnarray}
\label{WaveFn}
\Psi^{m\zeta}_\pm(R,\theta)=\frac{e^{im\theta}}{\sqrt{2}\Delta_m^{\zeta}}\begin{pmatrix}
\mp\big(m+\beta-\frac{\zeta}{2}\big) \cos\phi\, e^{-i\zeta\theta}\\
i\Delta_m^{\zeta}\\
\pm\big(m+\beta+\frac{\zeta}{2}\big) \sin\phi\, e^{i\zeta\theta}
\end{pmatrix}.
\end{eqnarray}
Additionally, we obtain the wave function associated with $E_{\rm FL}^{m\zeta}$ as
\begin{eqnarray}
\label{Wf1}
\Psi_{\rm Fl}^{m\zeta}(R,\theta)=\frac{ e^{im\theta}}{\Delta_m^\zeta}
\begin{pmatrix}
(m+\beta+\frac{\zeta}{2})\sin\phi\, e^{-i\zeta\theta}\\
0\\
(m+\beta-\frac{\zeta}{2})\cos\phi\, e^{i\zeta\theta}
\end{pmatrix}.
\end{eqnarray}
It is interesting to note that the flat band contains infinitely degenerate levels [see Eq. (\ref{Wf1})]; that is, any value of the quantum number $m$ yields a zero-energy solution.

\subsection{Discussion on the energy spectrum}\label{SecIIa}
\begin{figure}[h!]
\centering
\includegraphics[width=8.5cm, height=6cm]{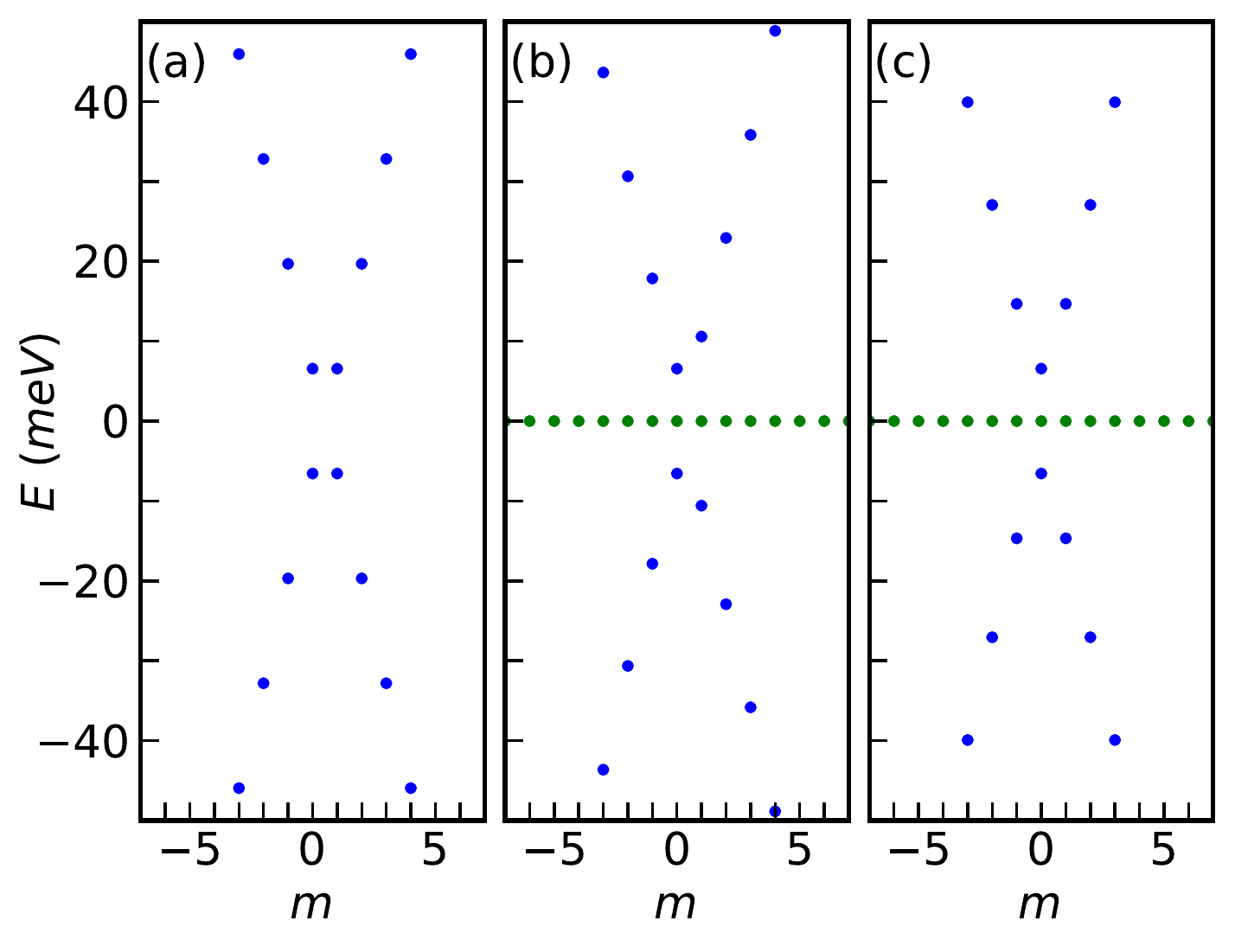}
\caption{Zero magnetic field energy levels of the $\alpha$-$T_3$ ring as a function of the quantum number $m$ for (a) $\alpha=0$, (b) $\alpha=0.5$, and (c) $\alpha=1$ in the $K$ valley. Here, we consider $R=50$ nm. The $\alpha=0$ case in (a) represents the quantum ring made of graphene, that is, without an atom sitting in the center of the hexagon. The flat band is missing in this case.}
\label{fig:ZeroF_Spect}
\end{figure}
The zero-field energy spectra in the $K$ valley for different values of $\alpha$ are shown in 
Fig. \ref{fig:ZeroF_Spect}.
One can easily verify the results of the graphene QR by setting 
$\alpha=0$ in Eqs. (\ref{Energ_Eig}) and (\ref{Delm}). In this case,
$\Delta_m^\zeta$ depends on the valley as $\Delta_m^\zeta=\vert m-\zeta/2\vert$. In the 
$K$ valley, the energy levels with $m=0$ and $m=1$ are degenerate, and so are levels with $m=2$ and $m=-1$, $m=3$ and $m=-2$, etc. Similarly, the energy levels with $m=0$ and $m=-1$, $m=1$ and $m=-2$, $m=2$ and $m=-3$, etc., are degenerate in the
$K^\prime$ valley. However, $\Delta_m^\zeta$ becomes valley independent, as $\Delta_m^\zeta=\sqrt{m^2+1/4}$, in the other limit corresponding to $\alpha=1$.
Therefore, the energy level with $m=0$ is nondegenerate, and the levels corresponding to $m=\pm1, \pm2,...$ are degenerate in both valleys.
At an intermediate value of $\alpha$ ($0<\alpha<1$), all the energy levels are nondegenerate. Here, only the $m=0$ level is valley independent while all the other levels depend on the valley in a complicated manner according to Eq. (\ref{Delm}).

\begin{figure}[h!]
\centering
\includegraphics[width=8.5cm, height=6cm]{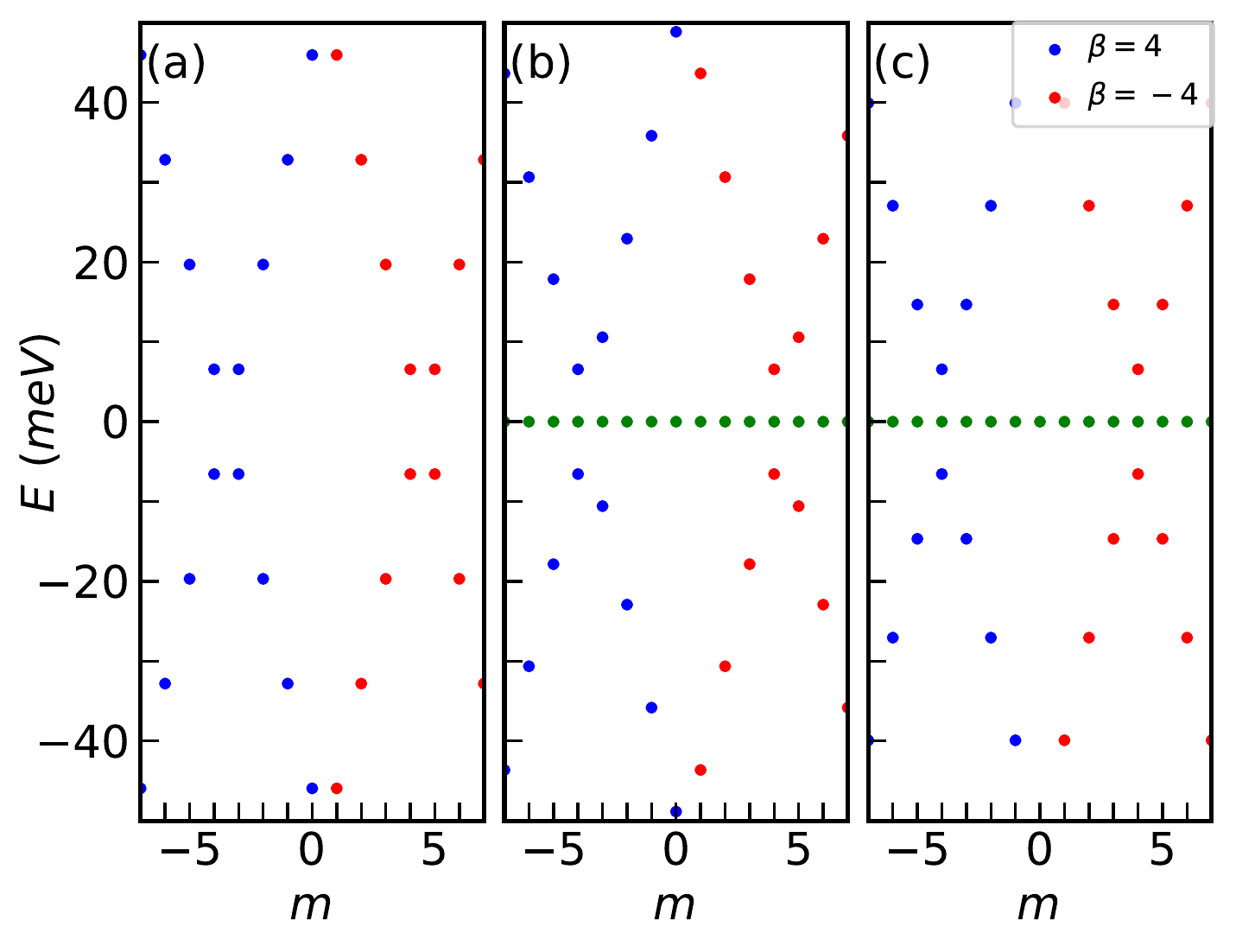}
\caption{The energy levels of the $\alpha$-$T_3$ ring in the 
$K$ valley as a function of the quantum number $m$ for (a) $\alpha=0$, (b) $\alpha=0.5$, and (c) $\alpha=1$ when the magnetic field is such that $\beta=\pm4$. Here, we consider $R=50$ nm. The $\alpha=0$ case in (a) represents the quantum ring made of graphene, that is, without an atom sitting in the center of the hexagon. Thus, the flat band is missing here.}
\label{fig:IntF_Spect}
\end{figure}
Now let us discuss the case in which the $\alpha$-$T_3$ ring is threaded by a perpendicular magnetic flux. In this case, we find 
$\Delta_m^\zeta=\vert m+\beta-\zeta/2\vert$ and $\Delta_m^\zeta=\sqrt{(m+\beta)^2+1/4}$ for $\alpha=0$ and $\alpha=1$, respectively. In the case of a graphene ring ($\alpha=0$), it is understood that the energy levels with $m=-\beta$ and $m=1-\beta$, $m=2-\beta$ and 
$m=-1-\beta$, $m=3-\beta$ and $m=-2-\beta$,etc., in the $K$ valley, and those with 
$m=-\beta$ and $m=-1-\beta$, $m=1-\beta$ and $m=-2-\beta$, $m=2-\beta$ and $m=-3-\beta$, etc., in the $K^\prime$ valley are degenerate when 
$\beta$ is an integer. In the other limiting case that is, $\alpha=1$, 
$\Delta_m^\zeta$ does not depend on the valley index $\zeta$.  Here, for integer values of $\beta$, the energy level with $m=-\beta$ is nondegenerate, and the energy levels with $m=\pm1-\beta, \pm2-\beta,...$ are degenerate. For intermediate values of $\alpha$, no such degeneracy exists. These facts are clearly shown in Fig. \ref{fig:IntF_Spect}. We have also verified (not shown here) that when $\beta$ deviates from an integer value, all the energy levels becomes nondegenerate for all values of $\alpha$.

\begin{figure}[h!]
\centering
\includegraphics[width=8.7cm, height=6.6cm]{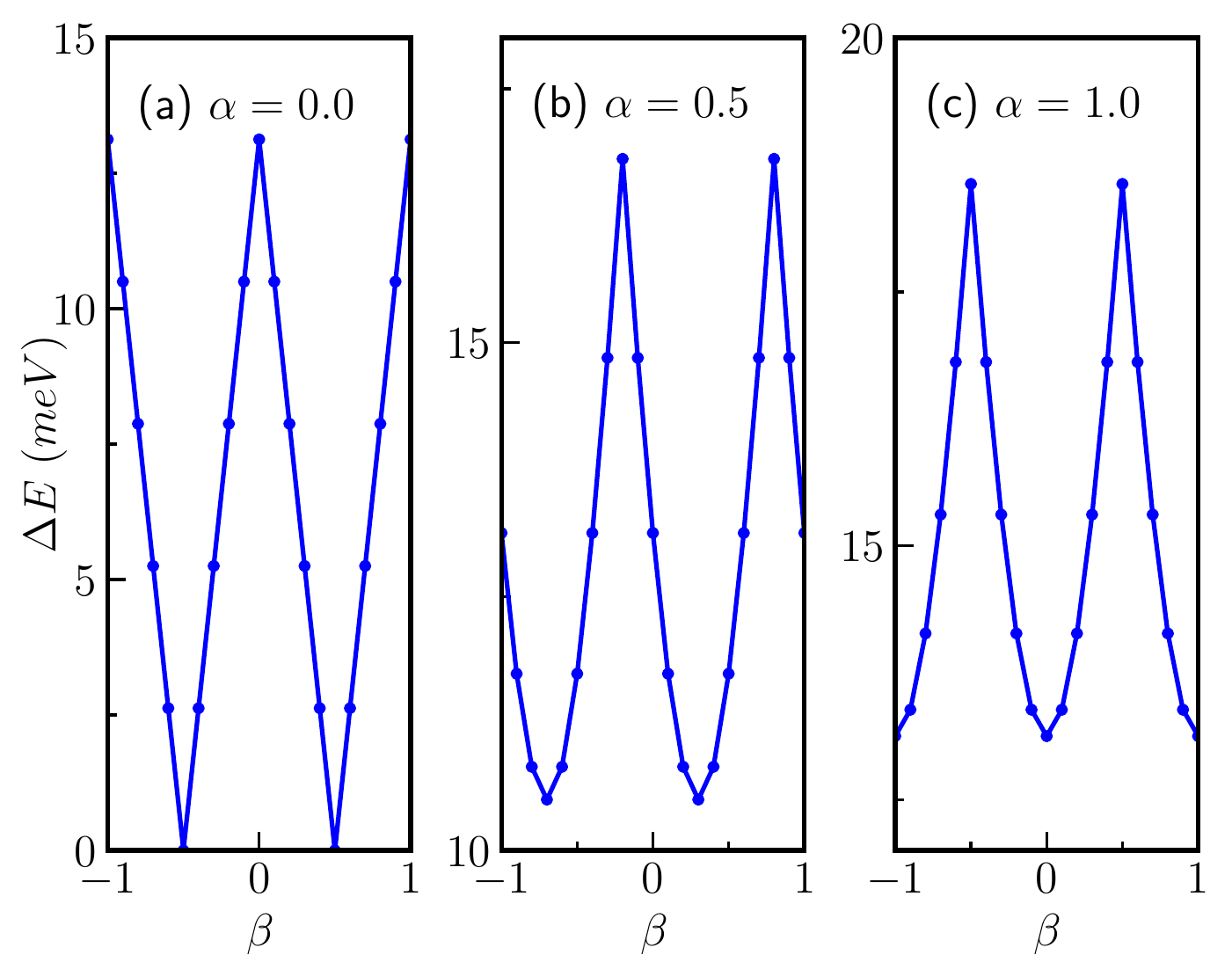}
\caption{The minimum value of the energy gap $\Delta E=E^m_+-E^m_-$ in the $K$ valley  as a function of $\beta$ for (a) $\alpha=0$, (b) $\alpha=0.5$, and (c) $\alpha=1$. Here, we consider $R=50$ nm.}
\label{fig:minE}
\end{figure}

Band gap tuning with magnetic field for this type of system can attract a lot of attention in absorption and emission spectra studies, where the lowest energy gap results in a strong signal in experiments. Motivated by this, let us define the energy gap in a particular valley $\zeta$ as, $\Delta E^\zeta=E_+^{m\zeta}-E_-^{m\zeta}$. It's minimum value in the $K$ valley is plotted as a function of
$\beta$ in Fig. \ref{fig:minE}. $\Delta E$ oscillates periodically with $\beta$. The period of oscillation is $\beta=1$. This is because the minimum magnetic field required to transfer an electron from one angular momentum state to the subsequent angular momentum state is such that $\beta=1$. For $\alpha=0$ and 
$\alpha=1$, $\Delta E$ is symmetric about $\beta=0$. However, $\Delta E$ is not symmetric about $\beta=0$ for an intermediate value, such as, $\alpha=0.5$. Note that the energy gap at $\beta=0$ (zero magnetic field) is independent of 
$\alpha$.

\begin{figure}[h!]
\centering
\includegraphics[width=9cm, height=7cm]{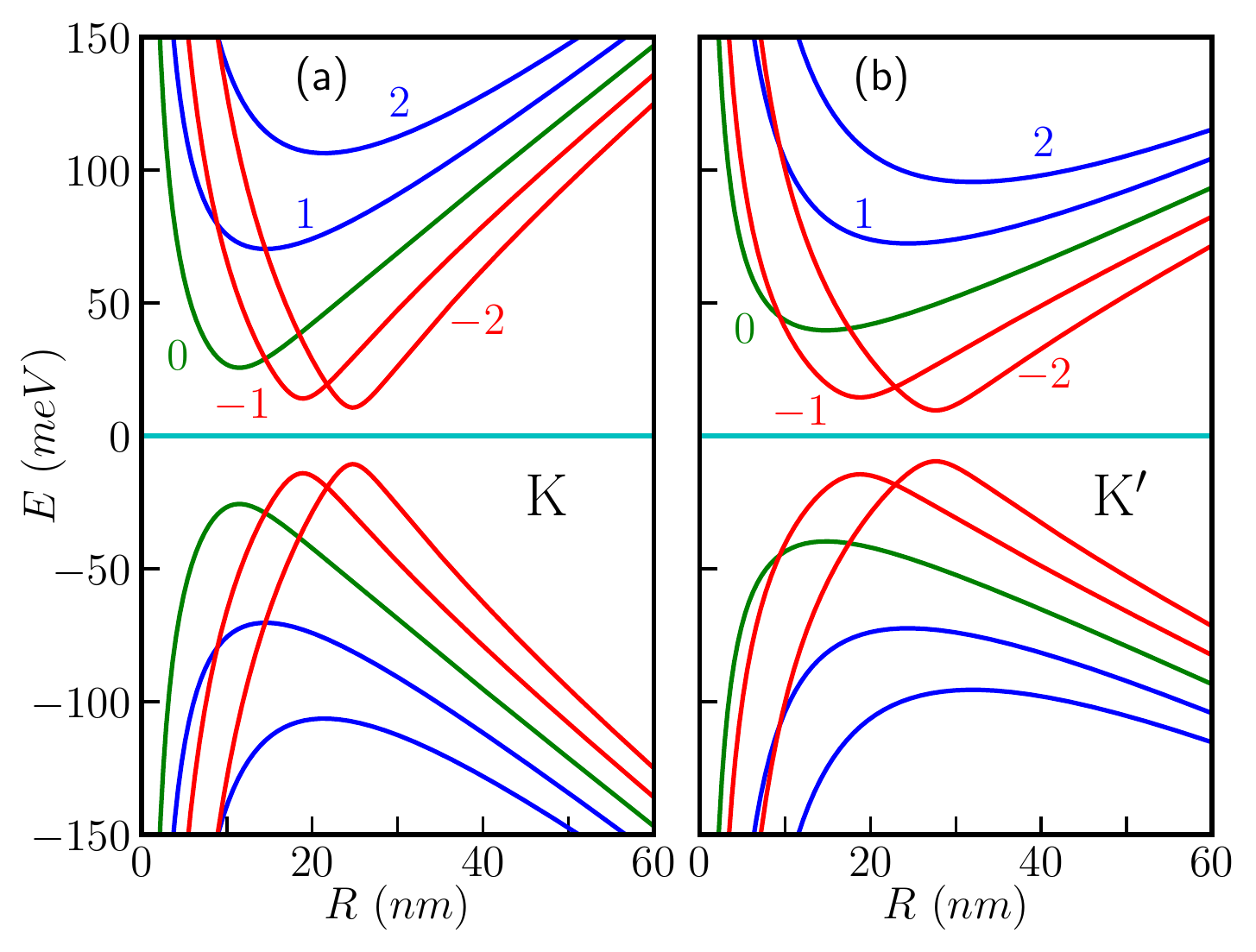}
\caption{Dependence of energy levels on the radius $R$ of the $\alpha$-$T_3$ ring are depicted for both valleys. We consider $B_0=5$ T and $\alpha=0.5$. Note that the valley degeneracy is lifted.}
\label{fig:EvsR}
\end{figure}

In order to ascertain the size dependence of the spectral properties, let us briefly discuss how the energy levels of the ring depend on the radius, $R$. It is clear from Eq. (\ref{Delm}) that $\Delta_m^\zeta$ is independent of $R$ in the absence of the magnetic field (since $R$ enters through the flux, $\Phi$ threads the ring) which leads to the $1/R$ dependence of the energy levels irrespective of the value of $\alpha$. The above scenario is altered significantly in the presence of a magnetic field. Figure \ref{fig:EvsR} shows the $R$ dependence of a few energy levels for both valleys considering
$B_0=5$ T and 
$\alpha=0.5$. Comparing Figs. \ref{fig:EvsR}(a) and \ref{fig:EvsR}(b) we conclude that all the energy levels become valley dependent in the presence of the magnetic field. Each of the levels show a nonmonotonic behaviour as a function of the radius $R$. In the limit of small $R$, all the energy levels vary inversely with $R$. On the other hand, the energy scales as $E\sim |ev_FB_0R|/2$ in the limit of large $R$. However, the criterion of $R$ being ``large" depends on $m$, which can be understood in the following way.

The energy level attains an extremum (the minimum for the conduction band and the maximum for the valence band) at a particular value of $R$, namely, $R=R_0$, which can be obtained by setting 
\begin{eqnarray}\label{En_ext}
\frac{dE_{\pm}^{m\zeta}}{dR}\Bigg\vert_{R=R_0}=0.
\end{eqnarray} 

\begin{figure}[h!]
\centering
\includegraphics[width=9cm, height=7cm]{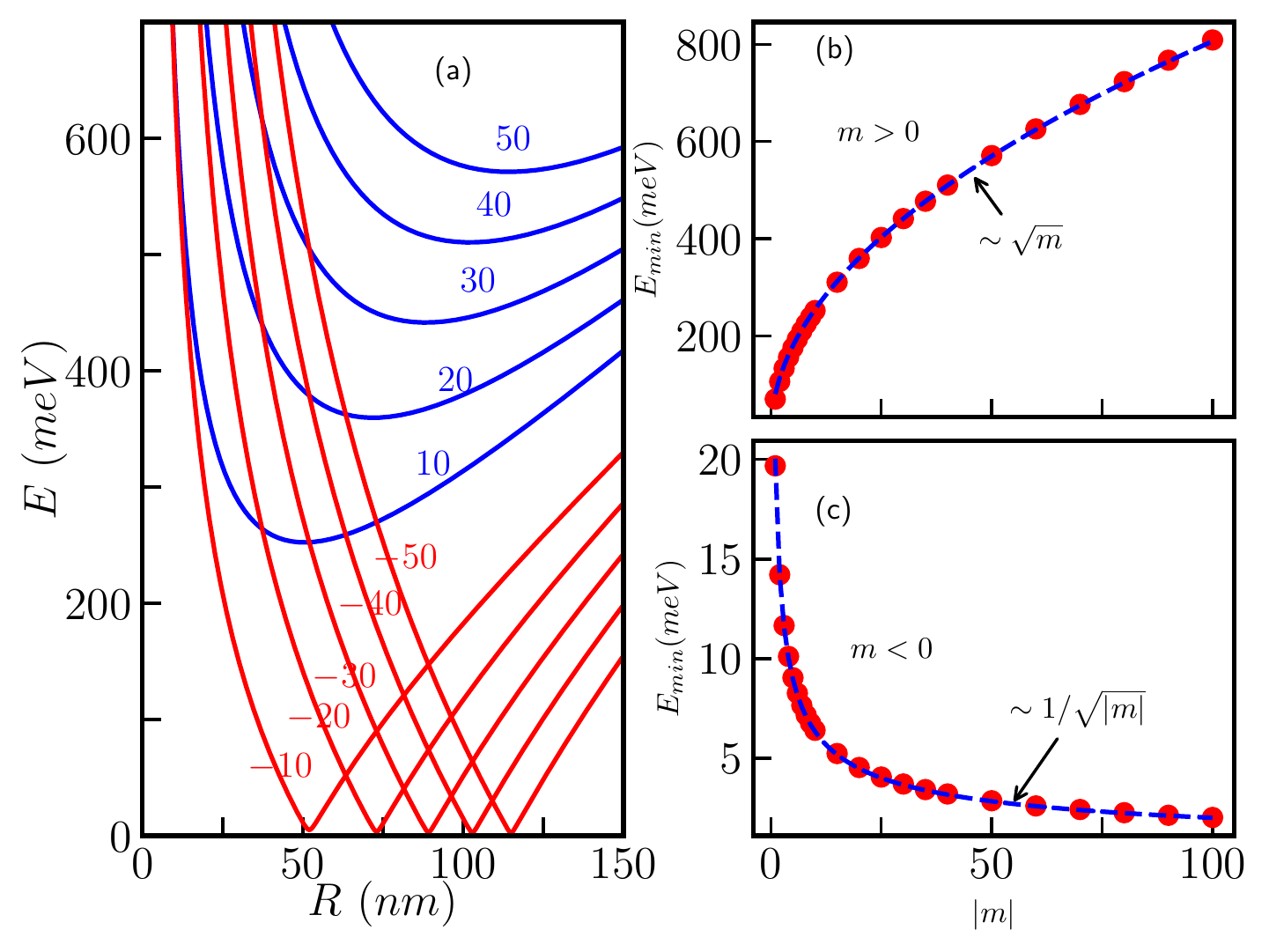}
\caption{(a) Dependence of energy levels on the radius $R$ of the $\alpha$-$T_3$ ring in the
$K$ valley considering large $m$ values. We consider $B_0=5$ T and $\alpha=0.5$. (b) Minimum value of the energy as a function of $m$ considering $m>0$. Here, we find $E_{min}\sim \sqrt{m}$. (c) Minimum value of the energy as a function of $m$ when $m<0$. In this case, we obtain $E_{min}\sim 1/\sqrt{\vert m\vert}$.}
\label{fig:EvsR_LargeM}
\end{figure}

Using Eqs. (\ref{Energ_Eig}) and (\ref{En_ext}), we find $R_0$ as
\begin{eqnarray}
R_0=\sqrt{2}l_0\Bigg(m^2-m\zeta\frac{1-\alpha^2}{1+\alpha^2}+\frac{1}{4}\Bigg)^{\frac{1}{4}},
\end{eqnarray}
where $l_0=\sqrt{\hbar/(eB_0)}$ is the magnetic length. For a fixed $B_0$, $R_0$ mainly depends on $m$, $\zeta$ and $\alpha$. In the two limiting cases, namely,
$\alpha=0$ and $\alpha=1$, we have $R_0=l_0\sqrt{2m-\zeta}$ and $R_0=\sqrt{2}l_0(m^2+1/4)^{1/4}$, respectively. For a given magnetic field, $R_0$ scales with $m$ as: $R_0\propto \sqrt{\vert m\vert}$ when $m$ is large enough. This feature is shown in Fig. \ref{fig:EvsR_LargeM}(a) where we present the radius dependence of a few positive energy levels with relatively large $m$, namely, $m=-50,-40,-30,-20,-10,10,20,30,40,50$ in the $K$ valley for $\alpha=0.5$ and $B_0=5$ T. It is clear that the condition for which one can consider $R$ to be ``large" depends on $m$ explicitly. More specifically, for an arbitrarily large value of $R$, we can always find an $m$ which corresponds to the energy minimum. The scaling of the minimum energy with $m$ also depends on the sign of $m$. For instance, $E_{min}\propto 1/\sqrt{|m|}$ when $m$ is negative. On the other hand, for positive $m$, we have $E_{min}\propto \sqrt{m}$. These scaling features of $E_{min}$
are depicted in Fig. \ref{fig:EvsR_LargeM}(b) and Fig. \ref{fig:EvsR_LargeM}(c). Therefore, the low-energy states in the $K$ valley are actually the states characterized by large negative $m$ values. It is also worth mentioning here that by reversing the sign of either the valley index $\zeta$ or the magnetic field $\bm B$, one can have $E_{min}\propto 1/\sqrt{m}$ for $m>0$ and 
$E_{min}\propto \sqrt{\vert m\vert}$ for $m<0$.

\subsection{Persistent Current}\label{P_cur}
\begin{figure}[h!]
\centering
\includegraphics[width=9cm, height=8cm]{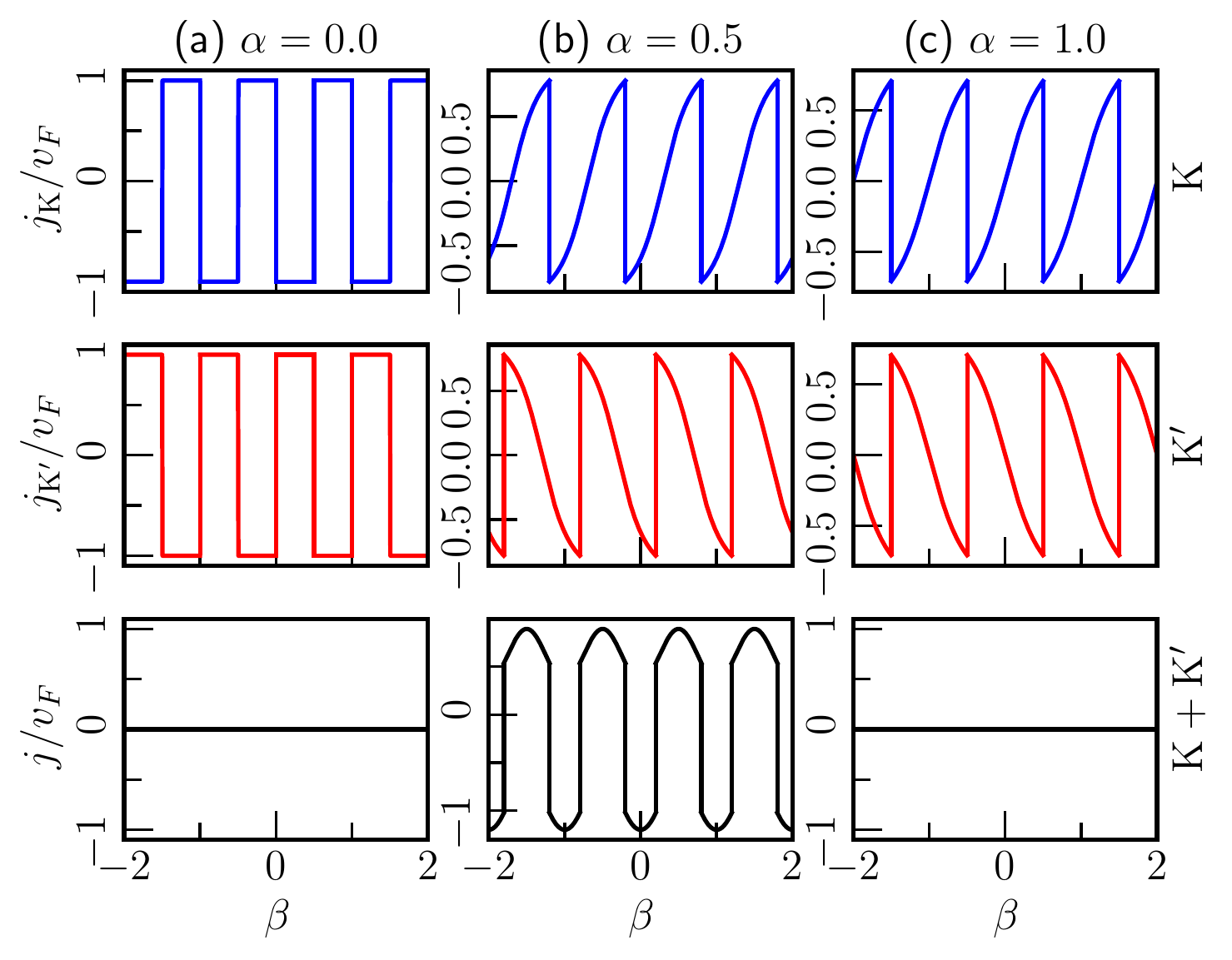}
\caption{Persistent current as a function of $\beta=\Phi/\Phi_0$ for (a) $\alpha=0$, (b) $\alpha=0.5$, and (c) $\alpha=1$. The first row is for the $K$ valley, the second row is for the $K^\prime$ valley, and the third one is for the total of the two valleys.}
\label{fig:Pers_Curr}
\end{figure}
Next, we discuss the behaviour of the persistent current in the $\alpha$-$T_3$ ring. The persistent current is the equilibrium current flowing along the angular direction in a QR when it is threaded by a magnetic flux. Proper knowledge of it aids in quantifying the energy spectrum near the Fermi energy. This current can be calculated using the relation $j_{x(y)}=v_F[\Psi^\dagger S_{x(y)}\Psi]$. Using this definition, the radial and the angular currents are further obtained as $j_r =v_F[\Psi^\dagger S_r \Psi]$ and 
$j_\theta =v_F[\Psi^\dagger S_\theta \Psi]$, respectively. Here, $S_r$ and $S_\theta$ are given by $S_r=S_x\cos\theta+S_y\sin\theta$ and $S_\theta =-S_x\sin\theta +S_y\cos\theta$.
Although the radial current vanishes,
we calculate the angular current in a particular valley as
\begin{eqnarray}\label{Persis_cur}
j_\zeta=\frac{v_F}{2\Delta_m^\zeta}\Bigg[2\zeta(m+\beta)-\frac{1-\alpha^2}{1+\alpha^2}\Bigg].
\end{eqnarray}

The total angular current is composed of the contributions from the individual valleys as  given by $j=j_\mathrm{K}+j_{\mathrm{K^\prime}}$. The expression for the persistent current given in Eq. (\ref{Persis_cur}) can also be obtained using the definition $j=-\sum_{m,\zeta}\frac{\partial E}{\partial \Phi}$ within the framework of the linear response theory, where the sum runs over all the occupied states.
In Fig. \ref{fig:Pers_Curr}, we show variation of the persistent current with $\beta$ for different values of $\alpha$. The persistent current is periodic in $\beta$ with a period of $\beta=1$. The oscillation pattern of the current corresponding to $\alpha=0$ is completely different than that for $\alpha=1$. Furthermore, the individual contributions arising from different valleys are exactly equal and opposite for both 
$\alpha=0$ and $\alpha=1$, which causes the total persistent current to vanish. 
The persistent current also vanishes in the individual valleys when $\beta=0$ for $\alpha=1$. The case for an intermediate $\alpha$ that is, $0<\alpha<1$ is more interesting. Here, the currents from the individual valleys do not compensate each other, which results in a nonvanishing total persistent current. It is also noteworthy that the current in a particular valley does not vanish even when $\beta$ is equal to zero, which gives rise to a total nonvanishing current corresponding to zero magnetic flux.

\section{Mass term}\label{Sec_Mass}
In this section, we are interested to see the effect of a mass term[\onlinecite{aT3_Gap1, aT3_Gap2}] on the low-energy spectrum of the $\alpha$-$T_3$ ring,
\begin{eqnarray}
M=\begin{pmatrix}
\delta & 0 & 0\\
0 & 0 & 0\\
0 & 0 & -\delta
\end{pmatrix}.
\end{eqnarray}
Note that $M$ can be thought to break the sublattice symmetry by including a different on-site potential in each of the $A$, $B$, and $C$ sublattices. With this, the effective Hamiltonian for the $\alpha$-$T_3$ ring in the presence of an external magnetic flux becomes,
\begin{widetext}
\begin{equation}
H^\zeta_\delta=\frac{\hbar v_F}{R}\begin{pmatrix}
\delta_0 & -i(m+\beta-\frac{\zeta}{2})\cos\phi & 0\\
i(m+\beta-\frac{\zeta}{2})\cos\phi & 0 & -i(m+\beta+\frac{\zeta}{2})\sin\phi\\
0 & i(m+\beta+\frac{\zeta}{2})\sin\phi & -\delta_0
\end{pmatrix},
\end{equation}
where $\delta_0=R\delta/(\hbar v_F)$.
The energy eigenvalues are obtained as,
\begin{equation}\label{Energy_mass}
E_k^{m \zeta} = 2\sqrt{\frac{P_m^\zeta}{3}}\cos\left[\frac{1}{3}\cos^{-1}\left(\frac{3Q_m^\zeta}{2P_m^\zeta}\sqrt{\frac{3}{P_m^\zeta}} \right)-\frac{2\pi k}{3} \right],
\end{equation}
where $k$= $0$, $1$, and $2$ are associated with the conduction band, the flat band, and the valence band, respectively. Here,
\begin{align*}
P_m^\zeta=\frac{\hbar^2 v_F^2}{R^2}\Big(\delta_0^2+{\Delta_m^\zeta}^2\Big),
\end{align*}
and
\begin{align*}
Q_m^\zeta=\delta\frac{\hbar^2v_F^2}{R^2}\Bigg[\frac{1-\alpha^2}{1+\alpha^2}\Big\{(m+\beta)^2+\frac{1}{4}\Big\}-\zeta(m+\beta)\Bigg].
\end{align*}
The normalized wavefunctions can be written in the form
\begin{equation*}
\Psi^{m\zeta}_k(R,\theta)=N_k^\zeta\, e^{im\theta}\begin{pmatrix}
-i\frac{\hbar v_F}{R}\big(m+\beta-\frac{\zeta}{2})(E_k^{m\zeta}+\delta) \cos\phi\, e^{-i\zeta\theta}\\
{E_k^{m\zeta}}^2-\delta^2\\
i\frac{\hbar v_F}{R}\big(m+\beta+\frac{\zeta}{2}\big)(E_k^{m\zeta}-\delta) \sin\phi\, e^{i\zeta\theta}
\end{pmatrix},
\end{equation*}
with
\begin{equation}\label{mass_WaveFn}
N_k^\zeta=\frac{1}{\sqrt{\frac{\hbar^2v_F^2}{R^2}[(E_k^{m\zeta}+\delta)^2\big(m+\beta-\frac{\zeta}{2})^2 \cos^2\phi+(E_k^{m\zeta}-\delta)^2\big(m+\beta+\frac{\zeta}{2}\big)^2 \sin^2\phi]+({E_k^{m\zeta}}^2-\delta^2)^2}}.
\end{equation}
\end{widetext}

\subsection{Discussion of the energy spectrum}
\begin{widetext}

\begin{figure}[!ht]
\centering
\includegraphics[width=15cm, height=10.2cm]{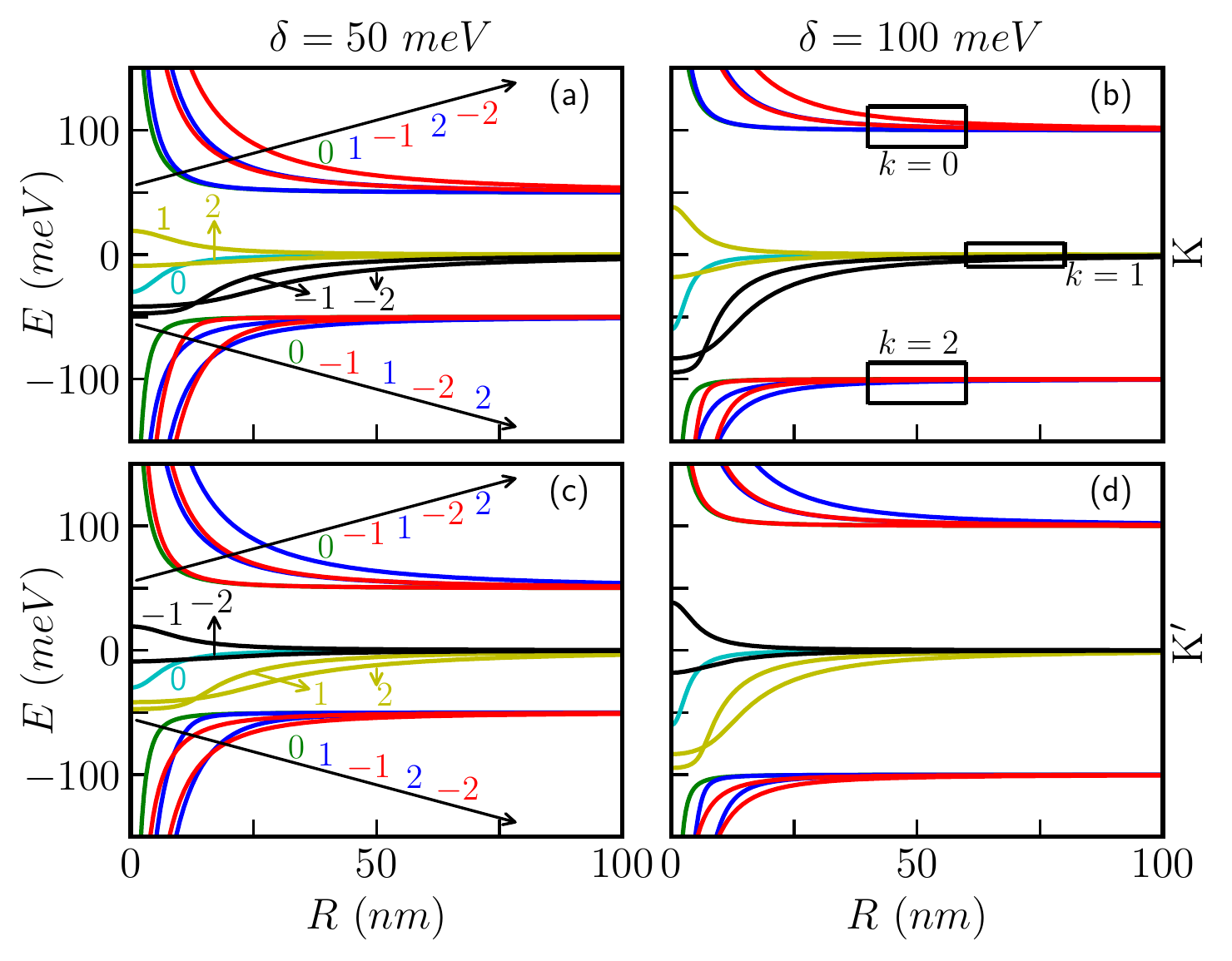}
\caption{Zero field energy levels as a function of radius $R$ for $\alpha=0.5$ at both valleys considering $\delta=50$ meV and $\delta=100$ meV.}
\label{fig:EvsR_Gap0}
\end{figure}
\begin{figure}[!ht]
\centering
\includegraphics[width=15cm, height=10.2cm]{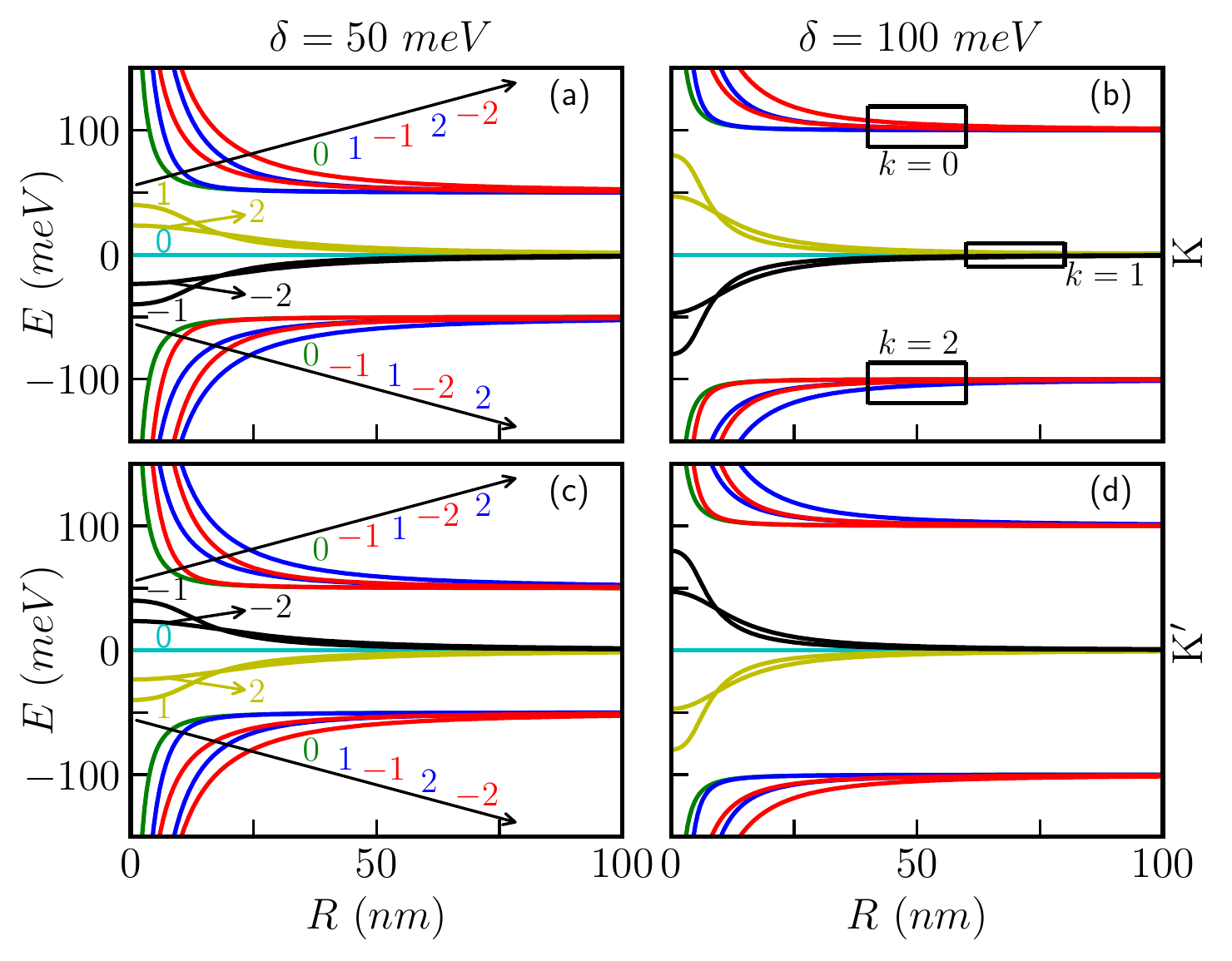}
\caption{Zero field energy levels as a function of radius $R$ for $\alpha=1$ at both valleys considering $\delta=50$ meV and $\delta=100$ meV.}
\label{fig:EvsR_Gap01}
\end{figure}
\end{widetext}
It is easy to confirm the results of Zarenia {\it et al.}[\onlinecite{Graph_Model1}] for 
graphene QR ($\alpha=0$) from Eq. (\ref{Energy_mass}). Figures \ref{fig:EvsR_Gap0} and \ref{fig:EvsR_Gap01} show the radius dependence of the zero-field energy levels at both valleys for 
$\alpha=0.5$ and $\alpha=1$, respectively. We consider two different values of the mass term, namely, 
$\delta=50$ meV and $\delta=100$ meV. The effect of the mass term on the energy spectrum is mainly two fold. First, it introduces gaps in the spectrum i.e., gap between the conduction band and the flat band and one between the flat band and the valence band. Second, it makes the flat band $(k=1)$ dispersive; that is, a nonzero group velocity is associated with each of the levels therein, which will contribute to the transport properties of the system. The energy levels with small $m$ values corresponding to $k=0$, $k=1$, and $k=2$ merge to $\delta$, $0$, and $-\delta$, respectively, in the limit of large $R$, as is evident from Figs. \ref{fig:EvsR_Gap0} and \ref{fig:EvsR_Gap01}. This merging of levels is also true for large $m$ in the case of $\alpha=1$, which we discuss in the next paragraph. In the small-$R$ limit, the energy levels belonging to $k=0$ and $k=2$ are inversely proportional to $R$ irrespective of
$\alpha$. However, the levels associated with $k=1$ deviate significantly from the $1/R$ dependence for both values of $\alpha$. 
For a given $k$ value, the energy levels in the $K$ valley are related to those in the $K^\prime$ valley as $E_k^{m(+\zeta)}=E_k^{-m(-\zeta)}$. Thus, the scenario is identical in the other valley, except $m$ reverses its sign. When $\alpha=1$ (Fig. \ref{fig:EvsR_Gap01}), we have 
$E_0^{m\zeta}=-E_2^{-m\zeta}$ in a particular valley $\zeta$, which is in direct contrast to the earlier result corresponding to $\delta=0$ (see Sec. \ref{SecIIa}). In the flat band ($k=1$), we find that the level with 
$m=0$ remains flat with zero energy, and the other levels satisfy $E_1^{m\zeta}=-E_1^{-m\zeta}$. In the case of $\alpha=0.5$ (Fig. \ref{fig:EvsR_Gap0}), the above-mentioned features are absent. In addition, the $m=0$ level in the flat band is no longer flat. It shifts down towards negative energy at small values of $R$. By inspecting both Figs. \ref{fig:EvsR_Gap0} and \ref{fig:EvsR_Gap01}, we can also note that all the energy levels are nondegenerate for all values of $\alpha$ except for $\alpha=0$[\onlinecite{Graph_Model1}], which can be attributed to the presence of the mass term $\delta$.

Here, we would like to point out some features of the zero-field energy spectrum in the limit of large $m$. In the case of $\alpha=1$, we have $\Delta_m^\zeta\sim m$,  $P_m^\zeta\sim \hbar^2 v_F^2 m^2/R^2$, and
$Q_m^\zeta\sim -\zeta \hbar^2 v_F^2\delta m/R^2$ when $m$ is large. Let us now define 
$$I_m^\zeta=\frac{3Q_m^\zeta}{2P_m^\zeta}\sqrt{\frac{3}{P_m^\zeta}}.$$ For large $m$, we obtain $I_m^\zeta\sim -3\sqrt{3}\zeta R\delta/(2\hbar v_F m^2)$. Considering typical values of the parameters, $R=200$ nm, 
$\delta=50$ meV, and $m=30$, we can verify that 
$3\sqrt{3}\zeta R\delta/(2\hbar v_F m)<1$. Therefore, we have $I_m^\zeta\to 0$ when $m$ is sufficiently large. From Eq. (\ref{Energy_mass}), we find the energy levels when $R$ is large enough 
\begin{eqnarray}\label{En_Largem}
E_{k}^{m\zeta}\sim \frac{2\delta}{\sqrt{3}}\cos\Big(\frac{\pi}{6}-\frac{2\pi k}{3}\Big).
\end{eqnarray}
It is now obvious from Eq. (\ref{En_Largem}) that $E_0^{m\zeta}\to \delta$, $E_1^{m\zeta}\to 0$, and $E_2^{m\zeta}\to -\delta$. In other words, the levels with large 
$m$ belonging to $k=0$, $k=1$, and $k=2$ also merge with $\delta$, $0$, and $-\delta$, respectively at large $R$ when $\alpha=1$. However, the scenario is different for $\alpha=0.5$. In this case, we obtain $I_m^\zeta\sim 9\sqrt{3}R\delta/(10\hbar v_F m)$ for large $m$. For the parameter values chosen earlier, we have $I_m^\zeta<1$. It is now straightforward to obtain the energy spectrum in limit of large $R$ as
\begin{eqnarray}
E_{k}^{m\zeta}\sim \frac{2\delta}{\sqrt{3}}\cos\Big(\theta_0-\frac{2\pi k}{3}\Big),
\end{eqnarray}
where $\theta_0$ depends on $R$, $\delta$, and $m$ explicitly.

\begin{widetext}

\begin{figure}[h!]
\centering
\includegraphics[width=13cm, height=8.5cm]{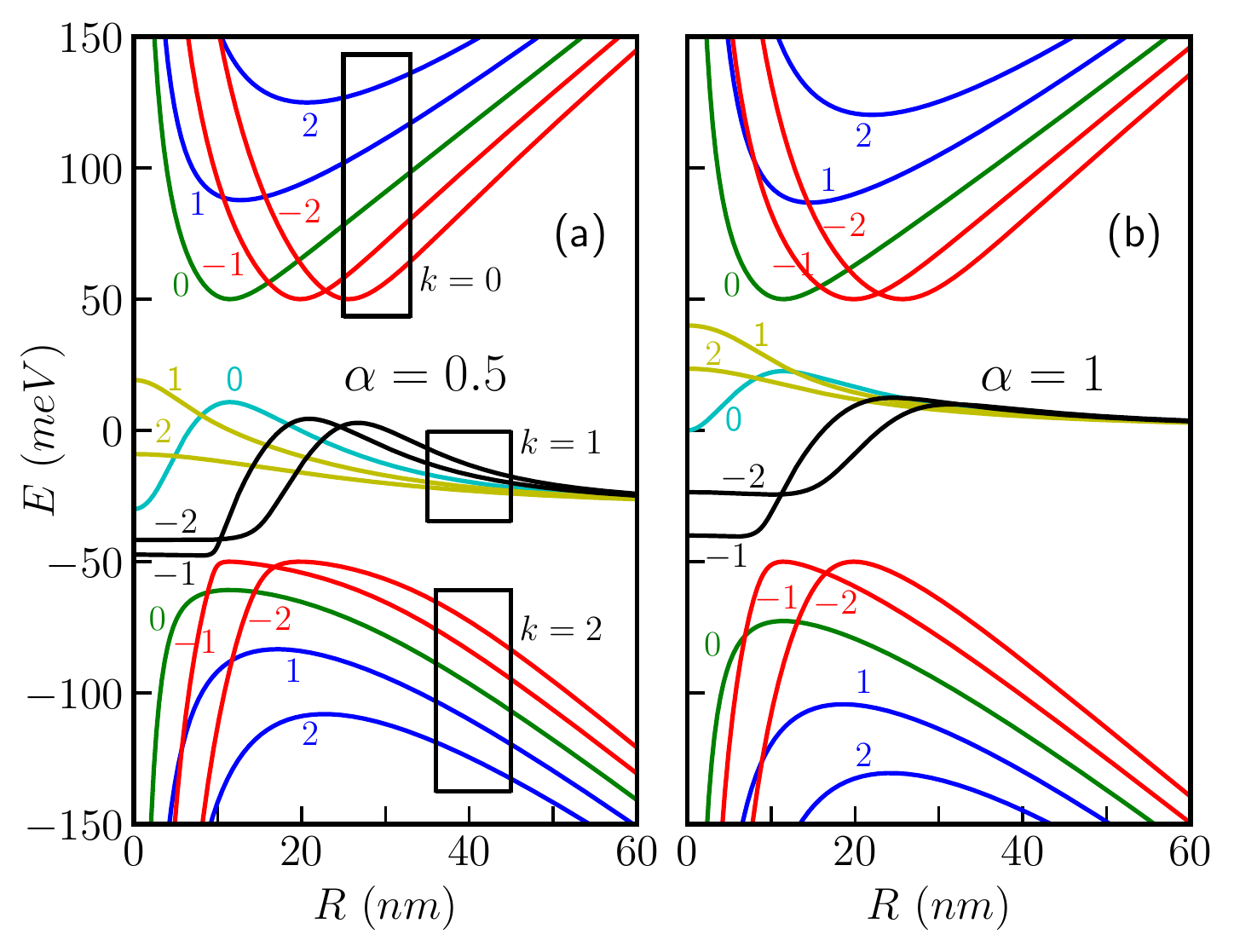}
\caption{Finite field energy levels in the $K$ valley as a function of $R$ in the presence of the mass term $\delta=50$ meV for (a) $\alpha=0.5$ and (b) $\alpha=1$. Here, we consider  $B_0=5$ T.}
\label{fig:EvsR_Gap5}
\end{figure}
\end{widetext}

In Fig. \ref{fig:EvsR_Gap5}, we show the $R$ dependence of the energy levels with small $m$ values in the $K$ valley in the presence of a magnetic field, where we have chosen, $B_0=5$ T. The qualitative features of the energy spectrum deviate significantly from the zero magnetic field case. Here, all the energy levels are non-degenerate for all values of $\alpha$. Unlike the $B_0=0$ case, we can observe that, $E_0^m \neq-E_2^{-m}$, $E_1^m\neq-E_1^{-m}$, and there is a distortion of the $m=0$ flat band energy level for $\alpha=1$ (Fig. \ref{fig:EvsR_Gap5}(b)). In addition, the levels in the $k=1$ band merge with zero energy at large values of $R$. However, this is not the scenario for intermediate values of $\alpha$, namely, $\alpha=0.5$ (Fig. \ref{fig:EvsR_Gap5}(a)), where the levels in the dispersive flat band do not merge to zero energy at large $R$. The energy levels in the conduction band and in the valence band depend on $R$ in a fashion similar to that in the case of $\delta=0$ (see Fig. \ref{fig:EvsR}).

\subsection{Persistent current}
Here, we also study the effect of the mass term on the variation of the persistent current. The persistent current in a particular valley and in a particular band now acquires the following form:
\begin{widetext}
\begin{eqnarray}
\label{P_cur_mass}
j_k^\zeta=2v_F\frac{\hbar v_F}{R}{N_k^\zeta}^2\big({E_k^{m\zeta}}^2-\delta^2\big)\Bigg[(m+\beta)E_k^{m\zeta}+\delta(m+\beta)\frac{1-\alpha^2}{1+\alpha^2}-\frac{\zeta\delta}{2}-\frac{\zeta E_k^{m\zeta}}{2}\frac{1-\alpha^2}{1+\alpha^2}\Bigg].
\end{eqnarray}
\end{widetext}

\begin{figure}[h!]
\centering
\includegraphics[width=9cm, height=8cm]{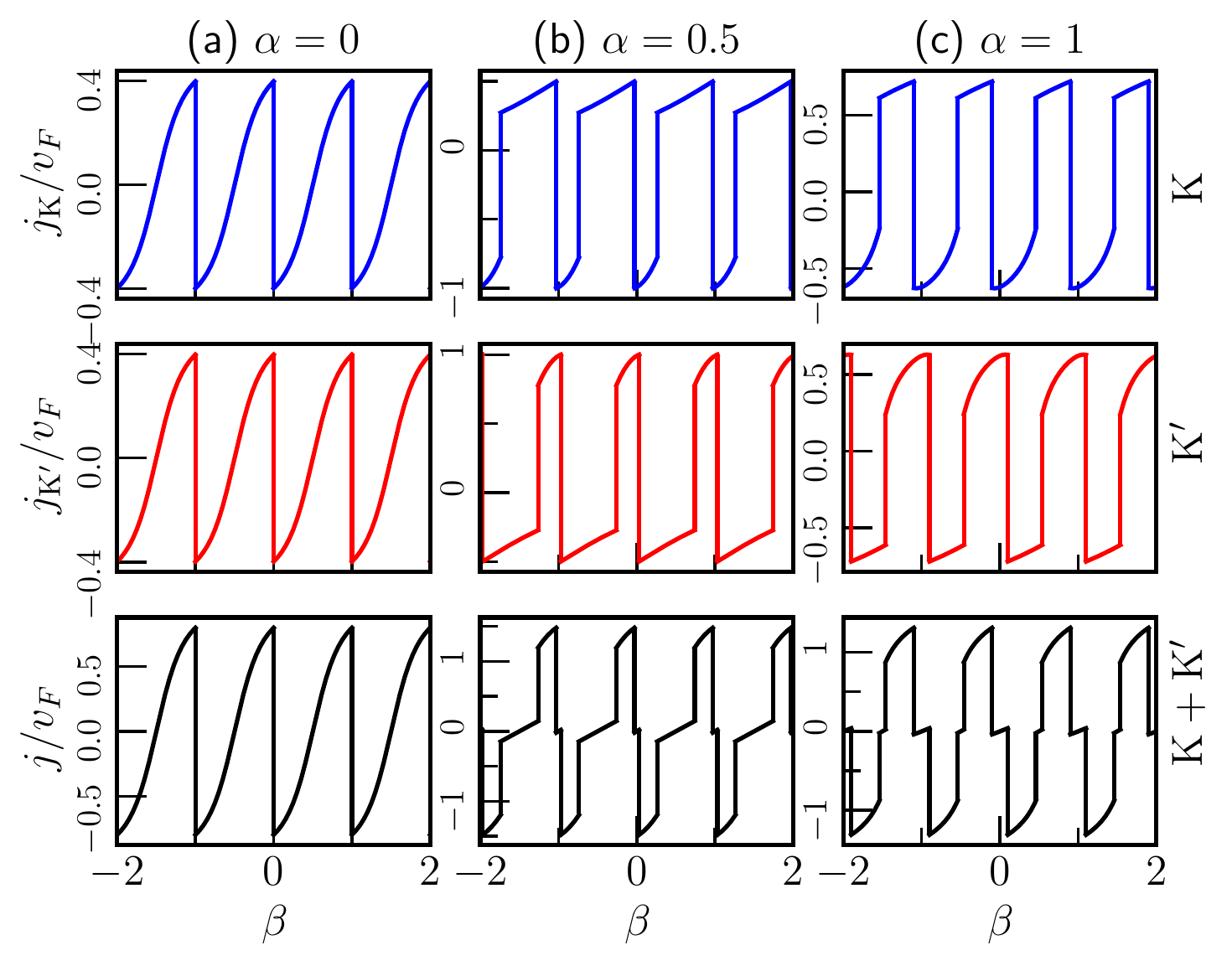}
\caption{Persistent current as a function of $\beta=\Phi/\Phi_0$ for (a) $\alpha=0$, (b) $\alpha=0.5$ and (c) $\alpha=1$. The first row is for the $K$ valley, the second row is for the $K^\prime$ valley, and the third one is for the total of the two valleys. Here, we have considered the mass term $\delta=50$ meV and the radius $R=10$ nm.}
\label{fig:Pers_Curr_mass}
\end{figure}

We can verify the results of the persistent current computed in Ref. [\onlinecite{Graph_Model1}] for graphene QR from Eq. (\ref{P_cur_mass}) by setting $\alpha=0$ and an appropriate value of $\delta$ considered there. The persistent current in a particular valley is calculated from the contributions from the conduction band ($k=0$) and distorted flat band $(k=1$) as $j_\zeta=j_\zeta^{k=0}+j_\zeta^{k=1}$. The total angular current comprising of contributions from both the valleys is given by $j=j_\mathrm{K}+j_\mathrm{K^\prime}$. It is worth mentioning that the distortion of the energy levels in the flat band gives rise to finite persistent current, unlike the case with no mass term ($\delta=0$). In Fig. \ref{fig:Pers_Curr_mass}, we show the variation of the persistent current with $\beta$ considering $\delta=50$ meV and $R=10$ nm. The introduction of the mass term completely changes the oscillation pattern of the persistent current from the case of $\delta=0$ (see Fig. \ref{fig:Pers_Curr}).
The currents at different valleys are no longer equal and opposite, which result in a nonvanishing total persistent current for all values of $\alpha$.
Further, the current in a particular valley and the total current oscillate periodically in $\beta$ with the periodicity $\beta=1$. Here, the total persistent current at 
$\beta=0$ (no magnetic field) is zero for all values of $\alpha \neq 0$. 

\section{Summary and outlook}\label{Sum}
In summary, we have investigated the electronic properties of the $\alpha$-$T_3$ ring analytically within a simple toy model. Particularly, we were interested in the behavior of the energy spectrum, the persistent current and the size dependences of the spectral features as one interpolates between graphene QR ($\alpha=0$) and the dice lattice QR ($\alpha=1$). Ignoring the radial dependence of the ring allows to overlook the boundary effects. We list a number of observations in the following. Confinement of the carriers in the ring leads to energy quantization characterized by the angular momentum quantum number $m$. As a result, we obtain discrete levels corresponding to both the conduction and in the valence bands. However, the flat band consists of a huge number of zero-energy degenerate levels which are insensitive to an applied magnetic field too. In the zero-field case, all the energy levels in the conduction or valence band depend inversely on the radius $R$ of the ring independent of values of $\alpha$. Furthermore, there is a degeneracy in the energy levels corresponding to $\alpha=0$. However, for an intermediate value of 
$\alpha$, namely, $0<\alpha<1$, the energy levels become nondegenerate. Interestingly, this degeneracy is restored in the case of $\alpha=1$ except for $m=0$ level. It is also found that the valley degeneracy is broken for all values of $\alpha$ such that $\alpha\neq 1$. When the ring is subjected to a perpendicular magnetic field, the energy levels follow a substantial deviation from their typical $1/R$ dependence. In the large $R$ limit the energy level scales as $E\sim R$, while at small $R$, it still behaves as $1/R$. The minimum energy gap between the conduction and the valence bands oscillates periodically as a function of the magnetic flux $\Phi$ with a period equal to one flux quantum $\Phi_0$. We also calculated the persistent current which exhibits $\Phi_0$ periodic oscillations in individual valleys, reminiscent of the Aharonov-Bohm oscillations. The total persistent current comprising both valley vanishes for both $\alpha=0$ and $\alpha=1$ as a consequence of exact compensation of the contributions from two valleys. But when $0<\alpha<1$, the total current is nonzero, and it undergoes $\Phi_0$-periodic oscillations. Interestingly, we observed a nonzero persistent current at zero magnetic field.
We also discussed the effect of an effective mass term on the energy spectra. In the absence of magnetic field, the mass term makes the flat band dispersive in the small-$R$ limit, except for the $m=0$ band corresponding to $\alpha=1$. A magnetic field alters the situation significantly by making all the levels in the flat band dispersive for all values of $\alpha\neq0$. With the mass term, the persistent current is again periodic with a period of one flux quantum, but the oscillation pattern is completely different from the previous case (zero mass). However, there is a finite contribution coming from the distorted flat band in the persistent current. Finally, the total current is nonzero for all values of $\alpha$. As a possible extension of our work, one could consider an $\alpha$-$T_3$ ring connected to external external leads in order to investigate the ballistic transport through the structure. In the presence of a perpendicular magnetic field, the conductance of the ring would exhibit quantum oscillations as the Fermi energy is varried. It would be interesting to see how these oscillations evolve over the entire range of $\alpha$.

\section*{ACKNOWLEDGMENTS}
T.B. sincerely thanks Prof. T. K. Ghosh for fruitful discussions. M.I. is also grateful to the Department of Physics, University of North Bengal, for providing local hospitality during his visit to pursue this work.

\section*{APPENDIX A: Justification of the substitution $\frac{\partial}{\partial r}\to -\frac{1}{2R}$}

The problem of the non-Hermiticity of the quantum ring Hamiltonian appears when the two-dimesional Hamiltonian contains a term linearly proportional to the momentum[\onlinecite{Sem_Ring1, Graph_Model1, Graph_Model2}]. To bypass such ambiguity, the substitution $\frac{\partial}{\partial r}\to -\frac{1}{2R}$ is necessary, as mentioned earlier[\onlinecite{Sem_Ring1, Graph_Model1, Graph_Model2}]. However, there is a nice argument in support of that substitution. It is based on the appropriate form of the radial momentum operator in two-dimensions.

Note that the operator $\mathcal{P}_r=\frac{\bm r\cdot \bm p}{r}$ is not Hermitian, i.e., 
$\mathcal{P}_r^\dagger\neq \mathcal{P}_r$. Therefore, the radial momentum operator can be written in the symmetric form 
\begin{eqnarray}
p_r = \frac{1}{2}(\mathcal{P}_r+\mathcal{P}_r^\dagger)
= \frac{1}{2}\Big(\frac{1}{r}\bm r\cdot \bm p+ \bm p\cdot \bm r\frac{1}{r}\Big).
\end{eqnarray}

Let us examine the action of $p_r$ on a differentiable function of $\bm r$, say, $g(\bm r)$. One can obtain
$$
p_r g(\bm r)=-\frac{i \hbar}{2}\Bigg[\frac{\partial g}{\partial r}+\bm \nabla \cdot \Big(\frac{\bm r g}{r}\Big)\Bigg]=-i \hbar\Bigg[\frac{\partial}{\partial r}+\frac{1}{2r}\Bigg]g(\bm r).
$$

Therefore, the radial momentum opertor in two dimensions is identified as 
\begin{eqnarray}
p_r=-i\hbar\Big(\frac{\partial}{\partial r}+\frac{1}{2r}\Big).
\end{eqnarray}
 
For the case of a 1D quantum ring, the radial motion is essentially frozen, that is, $p_r=0$. Therefore, the substitution $\frac{\partial}{\partial r}\to -\frac{1}{2R}$ is justified.

\end{document}